\documentclass[aps,10pt,pra,twocolumn,showpacs,showkeys,superscriptaddress]{revtex4-2}

\usepackage{hyperref}
\hypersetup{colorlinks=true, linkcolor=blue, filecolor=magenta, urlcolor=cyan,}
\usepackage{amsmath}
\usepackage{amsfonts}
\usepackage{amssymb}
\usepackage{graphicx}
\usepackage{bbold}

\begin{document}

\title{Parametric feedback cooling of librations of a nanodiamond in a Paul trap: Towards matter-wave interferometry with massive objects }
\author{Maria Muretova}
\affiliation{Ben-Gurion University of the Negev, Department of Physics and Ilse Katz Institute for Nanoscale Science and Technology, Be'er Sheva 84105, Israel}

\author{Yonathan Japha}
\affiliation{Ben-Gurion University of the Negev, Department of Physics and Ilse Katz Institute for Nanoscale Science and Technology, Be'er Sheva 84105, Israel}

\author{Marko Toro\v{s}}
\affiliation{ University of Ljubljana, Jadranska 19, SI-1000 Ljubljana, Slovenia}

\author{Ron Folman}
\affiliation{Ben-Gurion University of the Negev, Department of Physics and Ilse Katz Institute for Nanoscale Science and Technology, Be'er Sheva 84105, Israel}


\begin{abstract}
Quantum mechanics (QM) and General relativity (GR), also known as the theory of gravity, are the two pillars of modern physics. A matter-wave interferometer with a massive particle can test numerous fundamental ideas, including the spatial superposition principle -- a foundational concept in QM -- in completely new regimes, as well as the interface between QM and GR, e.g., testing the quantization of gravity. Consequently, there exists an intensive effort to realize such an interferometer. While several paths are being pursued, we focus on utilizing nanodiamonds (NDs) as our particle, and a spin embedded in the ND together with Stern-Gerlach forces, to achieve a closed loop in space-time.
There is a growing community of groups pursuing this path (see \cite{white_paper}).
We are posting this technical note (as part of a series of seven such notes) to highlight our plans and solutions concerning various challenges in this ambitious endeavor, hoping this will support this growing community.
Here, we present a theoretical study concerning the impact of rotations of the ND on the interferometric contrast and cooling of librations in a Paul trap. 
We have previously shown that for a first-generation Stern-Gerlach interferometer with an ND composed of $10^7$ atoms, it is sufficient to cool the center of mass to milli-Kelvin temperatures for achieving a spatial splitting on the order of nanometers. In this work, we similarly show that rotation does not have to be cooled to the ground state, and cooling to hundreds of rotational phonons is good enough.
Specifically, we describe and simulate parametric feedback cooling of librational modes of a charged ND levitated in a Paul trap. The cooling is performed by modulating the electric field of the trap in resonance with the libration frequencies, with feedback from measuring these librational motions and their phase. We examine the dependence of the efficiency of cooling on the electric potential and the shape of the object. 
We show that the required libration temperatures should be within reach in the very near future. We would be happy to make more details available upon request.
\end{abstract}

\maketitle

\newcommand{\be}{\begin{equation}}
\newcommand{\ee}{\end{equation}}
\newcommand{\vect}[1]{\left(\begin{array}{c} #1 \end{array}\right)}
\newcommand{\matthree}[1]{\left(\begin{array}{ccc} #1 \end{array}\right)}

\section{Introduction}

Quantum Mechanics (QM) is a pillar of modern physics. It dominates the sub-microscopic world of elementary particles, atoms and molecules together with light bodies such as photons or specific materials at very low temperatures, but there is no well-defined limit on its validity beyond these scales. It is thus imperative to test it in ever-growing regions of the relevant parameter space and in particular, examine the superposition principle on new scales\,\cite{Ball2021, Romero-Isart2011,Bassi2013}. The second pillar is 
general relativity (GR), also known as the theory of gravity, which is necessary for describing the evolution of the universe on the large scale of stars, galaxies and other massive bodies. Although these two fundamental theories are well-established, each one in its own scale, the unification of both of them has not yet been achieved and forms a crucial conceptual challenge. Hence, it is just as imperative to experimentally test the interface of these two pillars by conducting experiments in which foundational concepts of the two theories must work in concert. 
Experimentally, a significant clash between the two theories may be realized if one of the most fundamental aspects of QM -- the principle of superposition -- would be able to be implemented for massive bodies that generate a measurable gravitational field\,\cite{Bose2025}. 

The most advanced demonstrations of massive spatial superpositions have been achieved by Markus Arndt's group, reaching systems composed of approximately 2,000 atoms\,\cite{Fein2019, Brand2020,Shayeghi2020}. This will surely grow by one or two orders of magnitude in the near future, but the question is whether one can find a new technique that would push the state of the art much further in mass and spatial extent of the superposition.
Several paths are being pursued \cite{Pino2018,Kialka2022}, and we choose to utilize Stern-Gerlach (SG) forces.
The Stern-Gerlach interferometer (SGI) has proven in the last decade to be a very agile tool for atom interferometry\,\cite{Amit2019,Keil2021,Dobkowski2025}. Consequently, we, as well as others, aim to utilize it for interferometry with massive particles, specifically, nanodiamonds (NDs) with a single spin embedded in their center. Such a plan of action has been in the minds of many for some time now\,\cite{Scala2013,Wan2016,Margalit2021}.

Levitating, trapping, and cooling of massive particles, most probably a prerequisite for interferometry with such particles, has been making huge progress in recent years. Specifically, the field of levitodynamics is a fast-growing field \cite{Gonzalez-Ballestero2021}. 
Commonly used particles are silica spheres. As the state of the art spans a wide spectrum of techniques, achievements and applications, instead of referencing numerous works, we take, for the benefit of the reader, the unconventional step of simply mapping some of the principal investigators; these include Markus Aspelmeyer, Lukas Novotny, Peter Barker, Kiyotaka Aikawa, Romain Quidant, Francesco Marin, Hendrik Ulbricht and David Moore.  Relevant to this work, a rather new sub-field, which is now being developed deals with ND particles, where the significant difference is that a spin with long coherence times may be embedded in the ND. Such a spin, originating from a nitrogen-vacancy (NV) center, could enable the coherent splitting and recombination of the ND by utilizing Stern-Gerlach forces \cite{Wan2016,Scala2013,Margalit2021}. This endeavor includes principal investigators such as Tongcang Li, Gavin Morley, Gabriel H\'{e}tet, Tracy Northup, Brian D’Urso, Andrew Geraci, Jason Twamley and Gurudev Dutt.

The first goal of the ND SGI will be to test QM, and specifically the spatial superposition principle, in new regimes. Here the unique $T^3$ signal will provide a clear experimental signature of a coherent superposition\,\cite{Scala2013,Wan2016,Margalit2021}. We aim to start with an ND of $\sim 10^7$ atoms and extremely short interferometer durations. Closing a loop in space-time in a very short time is enabled by the magnetic gradients coming from the current-carrying wires of the atom chip\,\cite{Keil2016}. Such an interferometer will already enable us to test existing understanding concerning environmental decoherence (e.g., from blackbody radiation), and internal decoherence\,\cite{Henkel2024}, never tested on such a large object in a spatial superposition. As we slowly evolve to higher masses and longer durations (larger splitting), the ND SGI will enable the community to probe not only the superposition principle in completely new regimes, but in addition, it will enable testing specific aspects of exotic ideas such as the Continuous spontaneous localization hypothesis\,\cite{Adler2021,Gasbarri2021}. As the masses are increased, the ND SGI will be able to test hypotheses related to gravity, such as modifications to gravity at short ranges (also known as the fifth force), as one of the SGI paths may be brought in a controlled manner extremely close to a massive object\,\cite{Geraci2010,Geraci2015,Bobowski2024,Panda2024} (the chip itself or an object placed on it). Once SGI technology allows for even larger masses ($10^{11}$ atoms), we could test the Di\'{o}si-Penrose collapse hypothesis\,\cite{Penrose2014,Fuentes2018,Howl2019,Tomaz2024} and gravity self-interaction\,\cite{Hatifi2023,Grossardt2021, Aguiar2024} (e.g., the Schr\"{o}dinger-Newton equation). Here starts the regime of active masses, whereby not only the gravitation of Earth needs to be taken into account. Furthermore, it is claimed that placing two such SGIs in parallel will allow probing the quantum nature of gravity\,\cite{Bose2017,Marletto2017,Wan2016,Scala2013,Marshman2020SGI,Marshman2020L,Vicentini2024}. This will be enabled by ND SGI, as with $10^{11}$ atoms the gravitational interaction could be the strongest. Several publications\,\cite{vanDeKamp2020,Schut2023,Schut2024} have already calculated this, to show that one can find sweet spots for which the gravitational interaction is the strongest of all forces, so that any entanglement found between the two spins of the two NDs must indicate that the mediating force, namely gravity, is a quantized field. 

Let us emphasize that, although high accelerations may be obtained with multiple spins, we consider only a ND with a single spin as numerous spins will result in multiple trajectories and will smear the interferometer signal. Further down the road, multiple spins are also not adequate for the eventual experiment on the quantization of gravity as the observable of such a quantum-gravity experiment is entanglement between two spins, averaging over many spins smears the signal. We also note that working with an ND with less than $10^7$ atoms is probably not feasible because of two reasons. The first is that NVs that are closer to the surface than 20\,nm lose coherence, and the second is that at sizes smaller than 50\,nm, the fabrication errors become large, and a high-precision ND source becomes beyond reach\,\cite{Givon_ND_fabrication}. 

While an interferometer based on Stern-Gerlach splitting of atoms was successfully realized\,\cite{Amit2019,Keil2021,Dobkowski2025}, massive objects containing many atoms present a more challenging task. 
One of the important aspects of creating superpositions of massive objects and recombining them to obtain an interferometric signal is that they possess additional degrees of freedom (DoF) that may provide a which-way information that would destroy the interference signal, or in other words, lead to decoherence. One such example is the internal phonons that may be excited in the object during the interferometric sequence\,\cite{Henkel2024}. Another example is the rotational DoF of the object, whose dynamics may lead to distinguishability and decoherence\,\cite{Japha2023,Rusconi2022}. As a way to stabilize the alignment of the ND and decrease rotational effects on the interferometric coherence it was proposed to use gyroscopic 
stabilization by inducing a fast rotation of the object around the spin axis\,\cite{ZhouT2024,Wachter2025,Rizaldy2025}. This requires inducing a fast rotation of the object around a given axis, which was experimentally realized for a ND\,\cite{Jin2024}. However, even in this case, an initial cooling of rotation is required for a high-fidelity gyroscopic stabilization around the correct axis.

Furthermore, in order to enable the operation of an interferometer based on coherent Stern-Gerlach splitting of a ND it is necessary to cool the linear motion of the object -- namely, its translational DoF. 
The first reason why one needs to cool the CoM is in order to increase the coherence length of the particle. As explained in works dealing with the so-called Humpty-Dumpty effect \cite{Margalit2021}, in order to achieve a reasonable interference visibility, the coherence length must be larger than the experimental uncertainties in recombining the wavepackets, thus closing the loop in space-time. 
The second reason is that the magnetic gradient used for the SG splitting may be non-uniform, so that the phase depends on the initial position of the ND. For lowering the phase noise from shot to shot, we require that the variation of the initial position are small enough.
Our simulations show that for the initial (low mass, short duration) SGI, milli-Kelvin temperatures suffice for satisfying both requirements.

Ground-state cooling of levitated nano-objects has so far been demonstrated only for some specific cases. For CoM motion, state-of-the-art techniques have reached the ground state in one dimension \cite{Delic2020} and two dimensions \cite{Piotrowski2023}. By contrast, the best demonstrations of rotational cooling -- cavity cooling \cite{Pontin2023} and parametric feedback cooling \cite{Gao2024} -- have reduced the temperature only to the millikelvin regime, still several orders of magnitude above the rotational ground state. All of these achievements involved silica spheres held in optical tweezers; comparable results have yet to be obtained for NDs.
One of the most efficient ways to trap and cool the motion of levitated nano-objects is by their interaction with a beam of laser light, either in free space or in a cavity\,\cite{Bang2020,Rudolph2021,Kuhn2017,Stickler2016,Schafer2021,rademacher2025roto}. However, NDs are sensitive to continuous
illumination by laser beams and might be damaged
by intense light \,\cite{Rahman2016,Frangeskou2018,Jin2024}. In addition, internal heating
of the diamond may reduce the coherence time of
the NV center that holds the spin that is necessary for
SGI. At high temperatures, the spatial decoherence due to blackbody radiation also increases.
It is therefore necessary to develop a trapping and cooling mechanism based on non-optical forces\,\cite{Dania2021,Bykov2023}. Trapping and cooling of charged levitated NDs in ion traps was achieved in a few groups\,\cite{Delord2017,Delord2020,Frangeskou2018,Jin2024}, but significant improvement is still needed for taking this cooling down to the ground state. However, the required CoM milli-Kelvin temperatures are clearly within
reach.

Here we discuss trapping and cooling of rotational DoF of a charged ND in a Paul trap. Cooling of rotational DoF in a Paul trap was already proposed in a previous work\,\cite{Martinetz2021}, but it was not realized. 
We describe a possible mechanism based solely on feedback by electric fields. 
Parametric feedback cooling has been applied to optically levitated nano-particles\,\cite{Gieseler2012,Jain2016,Bang2020,Magrini2021}. For this task, a Kalman filter was implemented\,\cite{Setter2018,Ferialdi2019} and optimization algorithms are being continuously discussed\,\cite{Manikandan2023}.
Feedback cooling of nano-particles was usually performed by controlling the laser light intensities, but has not yet been done by using electric fields in the RF domain as in ion traps. 
We sketch such a dark cooling procedure and provide theoretical predictions for its efficiency as a function of the ND shape.
We show that the required rotational temperatures are within reach.

This technical note is part of a series of seven technical notes put on the archive towards the end of August 2025, including a wide range of required building blocks for the ND SGI. The structure of this report is as follows. In section\,\ref{sec:motivation} we present the motivation for the cooling task and give an estimation of the temperature of librational DoF that we should aim for. 
In sec.\,\ref{sec:EOM} we present the equations of motion for a ND in a Paul trap and specifically the dynamics of the rotational/librational modes. In particular, in subsection\,\ref{sec:secular} we develop the solutions for secular rotational motion of the object and present some examples for the secular librational frequencies for different ND shapes and the threshold temperature for librational dynamics. In sec.\,\ref{sec:cooling} we present the theoretical basis and numerical results for parametric feedback cooling and discuss the efficiency of the cooling as a function of the object's shape and the trap parameters. Based on the cooling rates and heating mechanisms such as collisions with background gas we provide estimations regarding the achievable temperature in the different libration modes, depending on the background gas pressure. 

\section{Motivation}
\label{sec:motivation}

The motivation of this work is to examine the way to cool the rotational DoF of a ND to a temperature that will allow the operation of a SGI. The force due to the interaction of the spin embedded in the ND with an external magnetic field gradient depends on the spin state but also on the alignment of the spin axis with respect to the magnetic field direction. It is therefore imperative to take care that the alignment of the ND is controlled during the interferometer sequence. Furthermore, the interaction of the spin with the magnetic field itself creates a torque that changes the alignment of the ND unless the spin is exactly aligned anti-parallel to the magnetic field. In addition to the influence of the ND alignment on the SG forces needed to form the interferometer, we have examined in our previous work\,\cite{Japha2023} the dynamics of the rotational DoF during the interferometer sequence and showed that the control of these dynamics by itself is crucial to the interferometric coherence. This is because these dynamics, which are different along the two interferometer paths, may lead to distinguishability of the two paths and hence to decoherence. It is therefore necessary to cool the rotational DoF to a low temperature in which these decoherence effects are minimal. 

Recently it was proposed to minimize the decoherence effects due to rotations by gyroscopic stabilization, which requires bringing the ND to a very fast rotation around its spin axis\,\cite{ZhouT2024,Wachter2025,Rizaldy2025}. This procedure raises some practical difficulties. In addition to the fidelity of stabilizing the correct axis as discussed previously, it requires a specific alignment of the ND that should coincide with a stable alignment in a diamagnetic potential created by the applied magnetic fields. In any case, we defer the discussion of the idea of gyroscopic stabilization in the context of the proposed SGI to future work and concentrate in this work on the most rudimentary interferometric scheme that is studied in Ref.\,\cite{Japha2023}, which is similar, in the context of the rotational DoF (but not in the context of SGI), to the configuration studied in Ref.\,\cite{Rusconi2022}. In this scheme the two interferometer states along the two paths are the state where the spin is anti-parallel to the quantization axis $|-\rangle$ and the magnetically insensitive state $|0\rangle$. 

\begin{figure}
\includegraphics[width=\columnwidth]{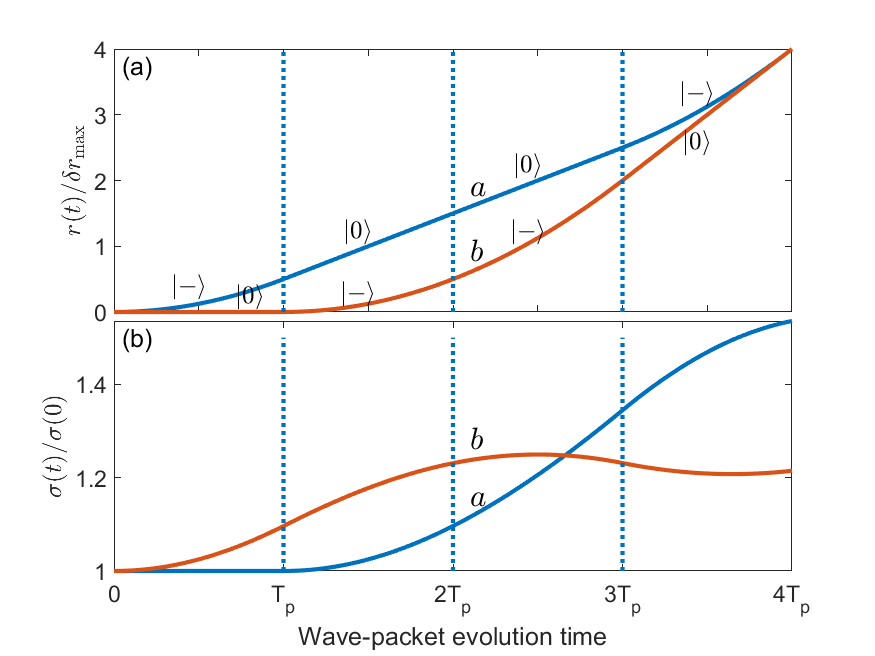}
\caption{Spatial and angular dynamics along the two interferometer paths (labeled $a$ and $b$). The spin states at the beginning of the two paths are the strong-field-seeking state $|-\rangle$ and the magnetically insensitive state $|0\rangle$ and they are flipped at $t=T_p$ and at $t=3T_p$. (a) Spatial trajectories: distance $r(t)$ from the initial point in units of the maximal distance $\delta r_{\rm max}$ between the paths. 
If the sequence is precise then both paths recombine at $t=4T_p$ with the same position and momentum, while the wavepacket spread is the same along the two paths.
(b) Angular evolution around a single rotation axis for the alignment prepared at the ground state of the magnetic potential, where the spin is anti-parallel to the field. 
While the expectation value of the spin direction remains anti-parallel to the magnetic field, its quantum uncertainty $\sigma $ is non-zero and diffuses the longitudinal nature of the spatial paths in (a). 
More importantly, the state-dependent angular potential [quadratic potential for the spin state $|-\rangle$] leads to a path-dependent evolution of the angular wave function (characterized by $\sigma$ and $\dot{\sigma}$). This prevents full recombination at the end of the sequence ($t=4T_p$). Taken from Ref.\,\cite{Japha2023}.}
\label{fig:interfscheme}
\end{figure}

Fig.\,\ref{fig:interfscheme} (taken from Ref.\,\cite{Japha2023}) demonstrates the spatial trajectories of the two interferometer paths and the evolution of the angular wave-packet widths of a single angular DoF (angle $\phi$ of the spin with respect to the magnetic field) along the two trajectories when the alignment of the ND is prepared such that the spin is aligned anti-parallel to the magnetic field with a quantum uncertainty $\Delta\phi=\sigma$. The coherence of the interferometer is determined by the overlap integral between the two wave packets at the output port of the interferometer. A thermal initial state can be described by a density matrix $\rho(\phi,\phi')=\sum_n w_n \psi_n(\phi)\psi_n^*(\phi')$, where $\psi_n(\phi)$ are eigenstates of the initial potential in which the angular state was prepared and $w_n\propto \exp(-E_n/k_BT)$ are the thermal weights. We assume that prior to the beginning of the interferometric sequence the spin state was $|-\rangle$ (with the spin oriented anti-parallel to the magnetic field) and that the angular state is in thermal equilibrium in a potential $V_-(\phi)=\mu\hat{\bf S}\cdot{\bf B}\approx -\mu |B|+\frac12 I\omega_B^2\phi^2$, where $\mu$ is the magnetic moment of the spin ($=h\times 2.8$\,MHz/G), $I$ is the moment of inertia around the given axis and $\omega_B=\sqrt{\mu|B|/I}$. Here we assume that the magnetic field is relatively weak, such that $\mu|B|$ is much smaller than the zero-field splitting ($D=h\times 2.87$\,GHz) between the state $|m_s=0\rangle$ and $|m_s=\pm 1\rangle$ of the NV center. The eigenstates in this harmonic potential are Hermite-Gaussian functions. 

The contrast of the interferometer due to the given angular dynamics is then
\be C=|{\rm Tr}[\rho \hat{U}_a^{\dag}\hat{U}_b]|=\left|\sum_n w_n\int d\phi\,\psi_n^{(a)}(\phi,t_f)\psi_n^{(b)\,*}(\phi,t_f)\right|\,, \ee
where $\hat{U}_a$ and $\hat{U}_b$ are the evolution operators along each of the paths and $\psi_n^{(a)}(t_f)$ and $\psi_n^{(b)}(t_f)$ are, respectively, the wave packets after evolution along the two paths for a time $t_f$ at the outport of the interferometer. 

Let us now examine the required temperature of the angular DoF that allows an interferometric contrast that will enable a measurable $T^3$ signal. We first consider a minimal interferometric operation where the measured phase will reach more than $4\pi$, such that at least two oscillations of the signal will be observed by scanning the interferometer time. 
We take a ND of mass $M\sim 5\cdot 10^{-19}$\,kg ($2.5\cdot 10^7$ atoms), so that it has an average diameter of $2(3M/4\pi\rho_D)^{1/3}\approx 65$\,nm ($\rho_D=3.51$\,gr/cm$^3$) and a moment of inertia $I\sim 2.1\cdot 10^{-34}$\,kg$\cdot$m$^2$. This mass enables the ND to have a shape with significantly different sizes along the different axes, as required for efficient cooling, while allowing the spin in the center to be located far enough from the surface (more than 20\,nm). 
the angular harmonic frequency for a magnetic field of 10\,G is $\omega_B=2\pi\times 1.5$\,kHz. The interferometer phase due to the spatial dynamics is given by 
\be \varphi_{\rm int}=\frac{M}{\hbar}a_-(a_-+2g_{\parallel})T_p^3\,, 
\label{eq:int_phase} \ee
w where $a_-$ is the acceleration of the state $|-\rangle$ and $g_{\parallel}$ is the gravity component along the direction of the magnetic gradient. Here we consider a scheme where the magnetic gradient is aligned with gravity and take an atom-chip configuration where the ND is initially placed at a distance of 2\,$\mu$m from the center of a wire carrying 1\,A on the chip. Another bias field generated by an external source or by parallel wires on the chip cancels the magnetic field at a point near the ND, such that the magnetic field at the position of the ND is about 10\,G and the magnetic gradient is $0.5$\,G/nm, giving rise to an acceleration $a_-=1.85$\,m/s$^2$ in the state $|-\rangle$. In this case, a quarter of the interferometer time $T_p=4\,\mu$s would give rise to an interferometer phase $\varphi_{\rm int}\approx 12$\,rad, and the maximal splitting is $a_-T_p^2=29.6$\,pm. 

\begin{figure}
\includegraphics[width=\columnwidth]{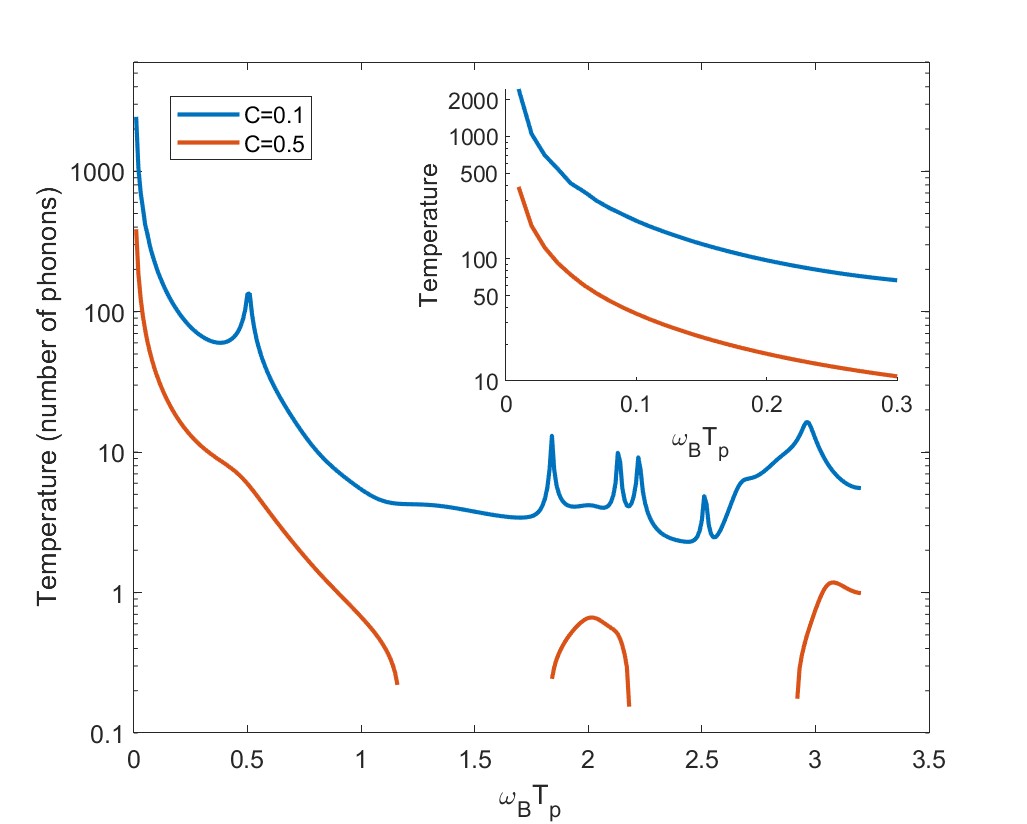}
\caption{Temperature (in units of a libration phonon energy divided by the Boltzmann constant) of a single angular degree-of-freedom (mode)  needed for achieving a given contrast of the interferometer signal as a function of $\omega_BT_p$, where $\omega_B=\sqrt{\mu|B|/I}$ is the libration frequency in the spin state $|-\rangle$ in a magnetic field $B$ and $T_p$ is the duration of each of the four interferometer gradient pulses. We assume that initially the libration mode is in thermal equilibrium in the harmonic torque due to the interaction of the spin in state $|-\rangle$ with the magnetic field. The interferometer sequence, which lasts for a time duration $4T_p$, is demonstrated in Fig.\,\ref{fig:interfscheme} (for the case where there is no gravity in the direction of the SG force). To obtain the temperature in Kelvin, the values in the plot should be multiplied by $\hbar\omega_B/k_B$. For the case where the magnetic field at the location of the spin is 10\,G ($\omega_B=2\pi\times 1.5$\,kHz) and the pulse duration is $T_p=4\,\mu$s, $\hbar\omega_B/k_B=72$\,nK and $\omega_BT_p=0.0375$ so that the required temperature is about 7\,$\mu$K ($\sim 100$ libration phonons) for achieving a contrast of C=0.5 and 40\,$\mu$K ($\sim 580$ phonons) in order to achieve a contrast of C=0.1. For a ND of mass $5\cdot 10^{-19}$\,kg located $2\,\mu$m from a chip wire of cross-section $1\,\mu$m\,$\times$\,$1\,\mu$m carrying 1\,A (current density of $10^8$\,A/cm$^2$) an interferometer in the direction of gravity with $T_p^{\rm max}=4\,\mu$s will 
accumulate a phase [see Eq.\,(\ref{eq:int_phase})] of about $4\pi$\,rad, allowing observation of two oscillations upon scanning $0<T_p<T_p^{\rm max}$.}
\label{fig:contrast_temperature}
\end{figure}
Fig.\,\ref{fig:contrast_temperature} shows the temperature (in units of $\hbar\omega_B/k_B$) required for achieving a given interferometer contrast as a function of $\omega_BT_p$. For our specific example above, we have $\omega_BT_p=0.0375$ and we need a temperature of 7\,$\mu$K to achieve a contrast of 0.5 and 40\,$\mu$K to achieve a contrast of 0.1 (see caption). In the range of small $\omega_BT_p$ the required temperature drops like $1/\omega_BT_p$, so that the required temperature decreases in proportion to the interferometer duration and stays almost independent of the magnetic field or moment of inertia around the given angle.

In this work we consider a procedure in which the ND is trapped and cooled in a Paul trap while it is electrically charged. After this cooling the ND is adiabatically transferred into an intermediate trap, which may be formed with an optical tweezer, neutralized and brought adiabatically into the location of the interferometer near the chip\,\cite{Liran_ND_neutralization}. If the transfer of the ND from the Paul trap to the final position near the chip is performed carefully enough, then the temperature of its translational and rotational DoF will stay close to the temperature that they reached during the cooling in the Paul trap. We should then aspire to cool the ND to the desired temperature for achieving interferometric contrast, and we shall examine this procedure of cooling in the Paul trap throughout the next sections. 
\section{Equations of motion in a Paul trap}
\label{sec:EOM}

In this section we derive the equations of motion for the rotational degrees-of-freedom (DoF) of a charged ND in a Paul trap. This derivation provides us with the secular frequencies of libration when the rotational dynamics is confined to small angles around the equilibrium alignment in the trap. We also obtain an estimate of the threshold temperature of each libration mode, allowing a confined oscillatory motion around the equilibrium rather than free rotation. An alternative derivation was previously presented in a work of another group\,\cite{Martinetz2021}, and here we concentrate on equations of motion in the frame of the object for a ND having a shape with reflection symmetry around the three principal axes. 
\begin{figure}
\includegraphics[width=\columnwidth]{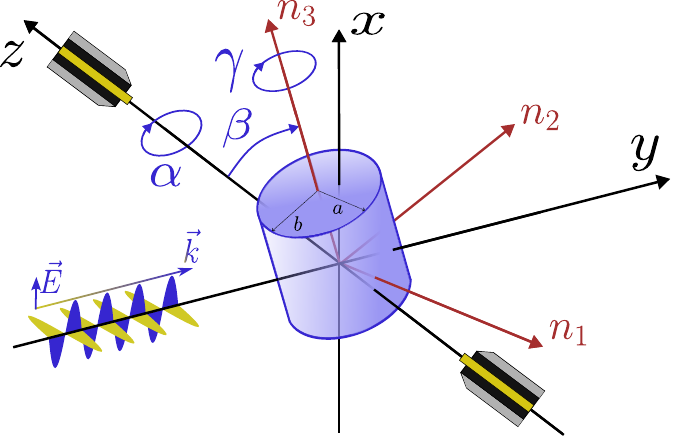}\label{fig:setup}
\caption{Illustration of an elliptical cylinder with short (long) semi-axis
denoted by $a$ ($b$) confined in a Paul trap. To describe the relative
orientation of the cylinder with respect to the laboratory reference
frame (i.e., the axis $x$, $y$, $z$) we introduce the body reference
frame (i.e., the axis $n_{1}$, $n_{2}$, $n_{3}$): the orientation
of the elliptical cylinder is specified by the Euler angles $\boldsymbol{\phi}=(\alpha,\beta,\gamma)$ in the $z$--$y'$--$z''$ convention. The symmetry axis of the electrodes is along the $z$-axis, which generates an equilibrium orientation given by $\boldsymbol{\phi}=(0,\pi/2,0)$ corresponding
to the body $n_{3}$ axis (i.e., the symmetry axis of the cylinder) aligned with the laboratory $x$-axis (i.e., along the vertical direction).
 The optical field, used for the detection, is shown to propagate in the positive direction of the $y$-axis and the polarization is assumed to be linear along the $x$-axis for concreteness.}
\end{figure}

\subsection{Basic equations}

We consider a ND with a total charge $q$ and an arbitrary shape in a Paul trap. The electric field is assumed to be a quadrupole field described by a potential 
\be V({\bf r})=\frac{U(t)}{2\ell_0^2}(\kappa_xx^2+\kappa_yy^2+\kappa_zz^2)\,, 
\label{eq:V} \ee
where $U(t)$ is a time-dependent voltage, $\ell_0$ is a characteristic length representing the distance between electrodes and $\boldsymbol{\kappa}=(\kappa_x,\kappa_y,\kappa_z)$ is a vector of elements on the order of 1 that satisfy $\sum_i\kappa_i=0$, such that the electric field derived from the potential ${\bf E}({\bf r})=-\nabla V({\bf r})$ satisfies Maxwell's equations. 

The above electric potential can be generated by various electrode configurations. The configuration that we will concentrate on when examining specific case studies will be based on two standard setups used in our lab, which are described in more detail in\,\cite{Feldman_Paul_trap_ND,Skakunenko_Needle_trap}. In these setups $\boldsymbol{\kappa}=(1-\epsilon,1+\epsilon,-2)^T$, where $\epsilon$ is an asymmetry  parameter with a value $\sim 0.05$. It follows that the object's equilibrium orientation is such that the long axis of the object is along the $x$ axis, where the confinement is weakest. As an example, we use a ND with a cylindroid shape as illustrated in Fig.\,\ref{fig:setup}, but the following derivation is valid for a general shape with reflection symmetry along the principal axes.

The force on a volume element of the object due to a charge distribution $\rho(x,y,z)$ is
\be d{\bf F}(x,y,z)={\bf E}(x,y,z)\rho(x,y,z)dx\,dy\,dz\,. \ee
We can expand the field around the center of mass ${\bf R}=(X,Y,Z)$ of the object, such that 
\be E_i(x,y,z)\approx E_i(X,Y,Z)+\sum_{j=1}^3 \frac{\partial E_i}{\partial r_j}(r_j-R_j)\,, \ee 
where the indices $i,j$ correspond to Cartesian axes and $r_j$ are the corresponding coordinates.
The force on the whole object is 
\be {\bf F}_{\rm CoM}=\int dx\,dy\,dz\,\rho(x,y,z){\bf E}({\bf r})
=q{\bf E}({\bf R})+({\bf p}\cdot\nabla){\bf E}\,, \ee
where $q=\int\rho\,dx\,dy\,dz$ is the total charge over the object and ${\bf p}=\int d^3{\bf r}\rho(x,y,z)({\bf r}-{\bf R})$ is the dipole moment due to non-symmetric charge distribution around the center of mass. The dipole term becomes significant compared to the force on the center-of-mass when the amplitude of oscillation of the object around the trap center is on the order of the displacement of the charge center from the mass center of the object, and may usually be neglected. 

Once the force on the CoM is known, it is a standard task to solve the equations of motion and derive the effective trapping frequencies due to the secular motion. We now concentrate on deriving the equations of motion for the angular (rotational) DoF.  
The alignment of the object in space can be described by the three Euler angles, which we denote by the vector $\boldsymbol{\phi}\equiv (\alpha,\beta,\gamma)^T$. The transformation between the coordinates ${\bf r}=(x,y,z)^T$ in the lab frame and the coordinates ${\bf r}'=(x',y',z')^T$ in the object's frame is represented by the rotation matrix $\hat{R}(\boldsymbol{\phi})$, such that
\be {\bf r}=\hat{R}(\boldsymbol{\phi}){\bf r}'\,. \ee
Here we use the convention $\hat{R}(\boldsymbol{\phi})=\hat{R}_z(\alpha)\hat{R}_y(\beta)\hat{R}_z(\gamma)$, where $\hat{R}_i(\theta)$ describes a rotation by an angle $\theta$ around axis $i$. The equations of motion for the Euler angles are given by
\be \dot{\boldsymbol{\phi}}=\hat{T}^{-1}\boldsymbol{\omega}\,, 
\label{eq:dotphi} \ee
where $\boldsymbol{\omega}=(\omega_1,\omega_2,\omega_3)^T$ are the angular frequencies around the principal axes, such that the angular momenta around these axes are $L_i=I_i\omega_i$, with $I_i$ being the moments of inertia. The matrix $\hat{T}^{-1}$ in Eq.\,(\ref{eq:dotphi}) is the inverse of the matrix
\be \hat{T}=\matthree{-\sin\beta\cos\gamma & \sin\gamma & 0 \\ \sin\beta\sin\gamma & \cos\gamma & 0 \\ \cos\beta & 0 & 1} 
\label{eq:T} \ee
With the elements related to the derivatives of the rotation matrix with respect to the Euler angles
\be \hat{T}_{ij}=-\frac12\epsilon_{ikl}\left(\hat{R}^T\frac{\partial \hat{R}}{\partial\phi_j}\right)_{kl}\,, 
\label{eq:Tij} \ee
with $\epsilon_{ikl}$ being the antisymmetric tensor and summation over $k\neq l\neq i$ is to be done. 
The equations of motion for the Euler angles in Eq.\,(\ref{eq:dotphi}) are complemented by the equations of motion for the angular velocities in the object's frame 
\be \dot{\omega}_i=\frac{1}{I_i}\left[\bar{\epsilon}_{ijk}(I_k-I_j)\omega_j\omega_k+\tau_i\right]\,, 
\label{eq:dotomega} \ee
where $\bar{\epsilon}_{ijk}=1$ only when $(ijk)$ is a symmetric permutation of $(123)$ and zero otherwise and $\boldsymbol{\tau}$ is the torque in the object's frame. The main effort in deriving the equations of motion for the rotational DoF is the calculation of the torque $\boldsymbol{\tau}$, which we will perform in the object's frame. 

\subsection{Calculation of the torques}

The torque $\boldsymbol{\tau} =\int d^3{\bf r}'\,({\bf r}'\times{\bf F}') $ can be obtained by expanding the electric field as 
\be {\bf E}({\bf r})={\bf E}({\bf R})+[({\bf r}-{\bf R})\cdot\nabla]{\bf E}\,. \ee
In the object's frame, the torque elements are
\be \tau_i=\epsilon_{ijk}[p_j E'_k({\bf R})+\sum_l Q_{jl}\hat{K}_{kl}]\,, \ee
where $p_j=\int dx'\,dy'\,dz'\,\rho(x',y',z')r'_j$, is the dipole moment in the object's frame, ${\bf E}'=\hat{R}^T{\bf E}({\bf R})$ is the field in the object's frame, and 
\be Q_{jl}=\int dx'\,dy'\,dz'\,\rho(x',y',z')r'_jr'_l\, \ee
is the quadrupole moment tensor, which is diagonal for objects with a reflection symmetry $Q_{jl}=Q_{jj}\delta_{jl}$. Here $\hat{K}$ is the matrix 
\be \hat{K}\equiv \hat{R}^T(\nabla\otimes {\bf E})\hat{R}\, 
\label{eq:K} \ee
is the field gradient tensor in the object's frame. As the field gradient is symmetric $\hat{K}_{ij}=\hat{K}_{ji}$ it follows that for an object with reflection symmetry
\be \tau_i=- \bar{\epsilon}_{ijk}(Q_j-Q_k)\hat{K}_{jk}\,, \ee 
Such that the torque is nonzero only if the object is rotationally non-symmetric about the corresponding axis of rotation. 
Specifically for the field derived from the potential in Eq.\,(\ref{eq:V}) the matrix $\hat{K}$ is given by
\be \hat{K}=-\frac{U(t)}{\ell_0^2}\hat{R}^T\hat{D}(\boldsymbol{\kappa})\hat{R}\,, \ee
where $\hat{D}(\boldsymbol{\kappa})$ is the diagonal matrix whose diagonal is the vector $\boldsymbol{\kappa}$.
It is worth noting that when the object is aligned with small angles with respect to the main axes of the trap, such that $\hat{K}$ is almost diagonal, the small deviation of $\hat{K}$ due to small angle deviations $\delta\boldsymbol{\phi}$ can be approximated by 
\be \hat{K}_{kl}\approx -\bar{\epsilon}_{kli}\hat{T}_{ii}(\hat{K}_{kk}-\hat{K}_{ll})\delta\phi_i\,, \ee
as following from Eq.\,(\ref{eq:Tij}), and we obtain the torque
\be \tau_i\approx \frac{U(t)}{\ell_0^2}\bar{\epsilon}_{ijk}(Q_j-Q_k)\hat{T}_{ii}(\kappa_j-\kappa_k)\delta\phi_i\,. \ee 

\subsection{Secular and micro-motion}

\begin{figure}
\includegraphics[width=\columnwidth]{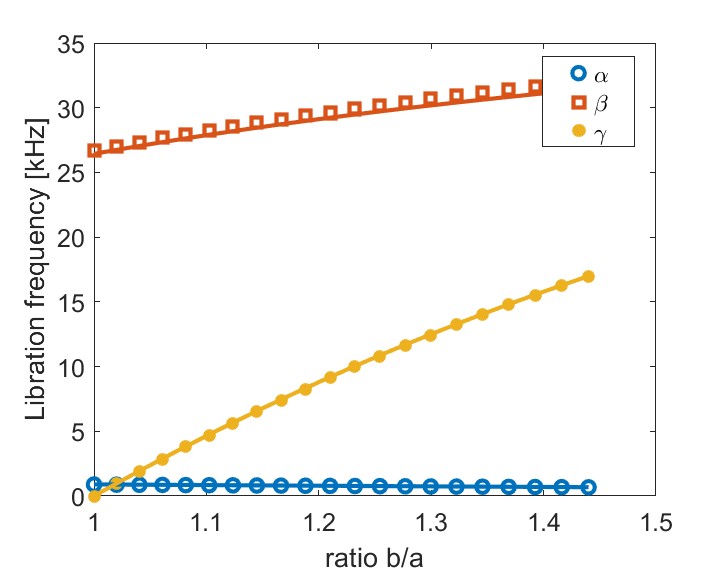}
\caption{Libration frequencies of a cylindroidal nano-object in a Paul trap as a function of a shape parameter: comparison of numerical calculation to analytical approximation [Eq.\,(\ref{eq:ws_approx})] for small libration amplitudes ($\delta\phi\approx 0.1$\,rad). Here we take a cylindroidal ND with length $L=100\,nm$ and basis semi-axes $a$ and $b$ such that $\sqrt{ab}=30$\,nm and $b/a$ is the x-axis of the plot. The ND has a total charge $q=100\,e$ and the uniform surface charge distribution. The Paul trap parameters are $U_{\rm AC}=100$\,V, $U_{\rm DC}=0$, $\ell_0=100\,\mu$m, $\boldsymbol{\kappa}=(-0.95,-1.05,2)$ and $\Omega_{\rm AC}=2\pi\times 250$\,kHz. With these parameters, the largest Mathieu parameter for the CoM motion is $q_z=0.52$ and for the librational motion it is $q_{\beta}=0.3$, giving rise to an excellent agreement between the analytical and numerical results. The numerical calculation is performed over 5\,ms and only one libration mode is taken to be excited.}
\label{fig:frequencies}
\end{figure}

\begin{figure}
\includegraphics[width=\columnwidth]{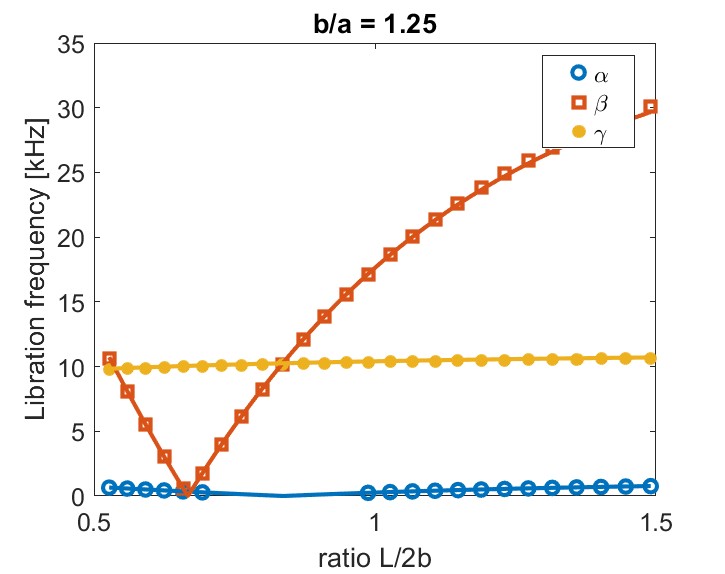}
\caption{Same as Fig.\,\ref{fig:frequencies} as a function of the ratio $L/2b$ with the ratio $b/a=1.25$ and the volume fixed. The rightmost point is the same as in Fig.\,\ref{fig:frequencies} with $L=100$\,nm and $\sqrt{ab}=30$\,nm, while the points at the left ($L/2b<1$) represent a shape where the width of the elliptic cross-section is larger than the height of the cylindroid. When $L/2b< 0.84$ the alignment of the object does not represent the minimal energy in the trap and the libration is meta-stable, with the frequency of libration around $\beta$ becoming zero at the transition point $L/2b\approx 0.84$. The numerical results (symbols) are not shown when the secular frequency is smaller than $2\pi/\tau_{\text{sample}}$, where $\tau_{\text{sample}}=5$\,ms is the sampling time duration.}  
\label{fig:frequencies_Lbratio}
\end{figure}

\label{sec:secular}

We now consider a potential with an oscillatory part
\be U(t)=U_{\rm DC}+U_{\rm AC}\cos\Omega_{\rm AC}t\,. \ee
If the oscillation frequency is large enough (as will be specified more explicitly below) the motion can be separated into micro-motion and secular (macro-) motion, such that the solutions for the Euler angles $\boldsymbol{\phi}=(\alpha,\beta,\gamma)$ can be written as
\be \boldsymbol{\phi}(t)=\boldsymbol{\phi}^{(s)}(t)+\boldsymbol{\phi}^{(m)}(t)\cos\Omega_{\rm AC}t\,. \ee
We can expand the torque $\boldsymbol{\tau}$ as
\be \boldsymbol{\tau}=\boldsymbol{\tau}(\boldsymbol{\phi}^{(s)})+(\boldsymbol{\phi}^{(m)}\cdot\nabla_{\phi})\boldsymbol{\tau}\, 
\label{eq:tau_expand} \ee 
and solve the equations of motion for the micro-motion at time scales of $\Omega_{\rm AC}^{-1}$ with the torque fixed at its value for the secular motion. Then we solve the equations of motion for the secular motion after averaging the torque over a period of the oscillation. For being able to do so, the elements of $\boldsymbol{\phi}^{(m)}$ must be very small relative to 1. 

The approximate solution for the equations of motion for the micro-motion is obtained by neglecting the time derivatives $\dot{\boldsymbol{\phi}}^{(s)}$ and terms that do not oscillate in the frequency $\Omega_{\rm AC}$. We then obtain 
\be \phi^{(m)}_i=-\frac{1}{\Omega_{\rm AC}^2}\hat{T}^{-1}_{ij}\frac{\tau^{\rm (AC)}_j(\boldsymbol{\phi}^{(s)})}{ I_j}\,, \ee
where $\hat{T}$ is given in Eq.\,(\ref{eq:T}) and in Eq.\,(\ref{eq:Tij}).
For calculating the second term in Eq.\,(\ref{eq:tau_expand}) we note that the torque for a symmetric object depends on the angles $\boldsymbol{\phi}$ through the term $\hat{K}=\hat{R}^T\hat{K}^{\text{LAB}}\hat{R}$, where $\hat{K}^{\text{LAB}}$ is the matrix of the vector field derivatives of Eq.\,(\ref{eq:K}). 
The derivative of this term with respect to the Euler angles $\boldsymbol{\phi}$ is given by 
\be \frac{\partial \hat{K}}{\partial\phi_i}=
\left(\hat{R}^T\frac{\partial\hat{R}}{\partial\phi_i}\right)^T\hat{K}+\hat{K}\left(\hat{R}^T\frac{\partial\hat{R}}{\partial\phi_i}\right)\,. 
\label{eq:dAdphi} \ee
By using Eq.\,(\ref{eq:Tij}) we obtain
\begin{eqnarray} 
\sum_j \frac{\partial\tau_i}{\partial\phi_j}\phi_j^{(m)} & = & -\epsilon_{ikl}Q_k \times 
\label{eq:tau_expand1} \\
&& \times \sum_j[-\epsilon_{lmn} \hat{K}_{mk}+\epsilon_{mkn}\hat{K}_{lm}]T_{nj}\phi_j^{(m)}\,. \nonumber
\end{eqnarray}
This can be written more explicitly as
\begin{eqnarray} 
\sum_j & \frac{\partial\tau_i}{\partial\phi_j}\phi_j^{(m)}
= -\frac{1}{2\Omega_{\rm AC}^2}\epsilon_{ikl}(Q_k-Q_l) \times \nonumber  \\ 
 & \times \left[\hat{K}_{ik}\frac{\tau_k^{\rm AC}}{I_k}-\hat{K}_{li}\frac{\tau_l^{\rm AC}}{I_l} 
-(\hat{K}_{kk}-\hat{K}_{ll})\frac{\tau_i^{\rm AC}}{I_i}\right]\,. 
\label{eq:tau_expand2}
\end{eqnarray}

When substituting for the torque $\tau_i^{\rm AC}=\epsilon_{imn}Q_m\hat{K}_{nm}=\frac12\epsilon_{imn}(Q_m-Q_n)\hat{K}_{mn}$ and averaging over a period of oscillation we obtain
\begin{eqnarray} \tau_i^{(s)} &=& \tau_i^{\rm DC}
-\frac{1}{4\Omega_{\rm AC}^2}\epsilon_{ijk}(Q_j-Q_k)\times \nonumber \\ 
&& \times \left[\hat{K}_{ij}\hat{K}_{ik}\left(\frac{Q_k-Q_i}{I_j}-\frac{Q_i-Q_j}{I_k}\right)\right. - \nonumber \\
&&\left. -(\hat{K}_{jj}-\hat{K}_{kk})\hat{K}_{jk}\frac{Q_j-Q_k}{I_i}\right]\, 
\label{eq:tauis} 
\end{eqnarray}
Now let us assume that the motion is around small Euler angle deviation from the main axes of a quadrupole field, in which $\partial_iE_j=\partial_iE_i\delta_{ij}$. Then the elements of the matrix $\hat{K}$ can be approximated by 
\be \hat{K}_{ij}\approx \hat{K}_{ii}\delta_{ij}-\sum_l \bar{\epsilon}_{ijk}T_{kl}(\hat{K}_{ii}-\hat{K}_{jj})\delta\phi_l\,. \ee
Then the first term in the square brackets in Eq.\,(\ref{eq:tauis}) is second-order in the small angular deviations and to first order we are left with the approximate expression
\be \tau_i^{(s)}=\tau_i^{\rm DC}
-\frac{1}{4I_i\Omega_{\rm AC}^2}\epsilon_{ijk}\hat{T}_{il}(Q_j-Q_k)^2(\hat{K}_{jj}-\hat{K}_{kk})^2\delta\phi_l\,. 
\label{eq:tauis_small} \ee
For small angles relative to the equilibrium orientation of the object, which is taken here to be aligned with the Cartesian axes if the angular DoF are trapped, the field tensor in the object's frame  $\hat{K}$ is diagonal and the elements of the matrix $\hat{T}$ are either $0$ or $\pm 1$, as it represents rotations by products of $\pi/2$.  It follows that the simple equations of motion are
\be \ddot{\phi}_i\approx \hat{T}^{-1}_{ij}\frac{\tau_j^{\rm DC}}{I_j}-(\Omega_i^{(s)})^2\phi_i\,, \ee
where the secular frequency of oscillation is
\be \Omega_i^{(s)}=|\hat{T}_{il}|\bar{\epsilon}_{ljk}\frac{|(Q_j-Q_k)(\hat{K}_{jj}-\hat{K}_{kk})|}{\sqrt{2}I_l\Omega_{\rm AC}}\,. 
\label{eq:ws_approx} \ee
This corresponds to effective Mathieu parameters for rotations
\be q_i=2\frac{U_{\rm ACa}}{\ell_0^2}|\hat{T}_{il}|\bar{\epsilon}_{ljk}\frac{(Q_j-Q_k)(\kappa_j-\kappa_k)}{I_l\Omega_{\rm AC}^2}\,. 
\label{eq:q} \ee
As the quadrupole moments $\mathbf{Q}$ are proportional to the charge and the moments of inertia $I_i$ are proportional to the mass, it follows that the mathieu coefficients and secular frequencies for librations have the same dependence on the trap properties as the corresponding Mathieu parameters and secular frequencies of the translational dynamics in the trap, except for the dimensionless geometric factor $|\hat{T}_{il}|\bar{\epsilon}_{ljk}[(Q_j-Q_k)/q](M/I_l)$ that depends on the object's shape and the ratio $(\kappa_j-\kappa_k)/\sqrt{|\kappa_j\kappa_k|}$ that depends on the orientation of the object relative to the axes of the quadrupole field. 

\begin{figure}
\includegraphics[width=\columnwidth]{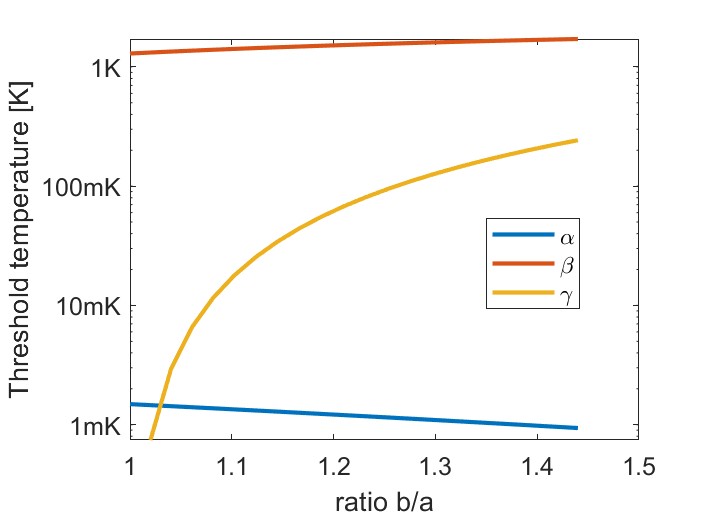}
\caption{Threshold temperatures [Eq.\,(\ref{eq:T_lib})] for libration dynamics for the three Euler angles of a ND in a Paul trap with parameters as in Fig.\,\ref{fig:frequencies}. For a cylindrical object ($b/a=1$) the rotation mode related to the Euler angle $\gamma$ (rotation around the cylinder axis) has no librational mode, and the threshold temperature for this mode grows with the asymmetry between the axes.} 
\label{fig:Tth_vs_r}
\end{figure}

\begin{figure}
\includegraphics[width=\columnwidth]{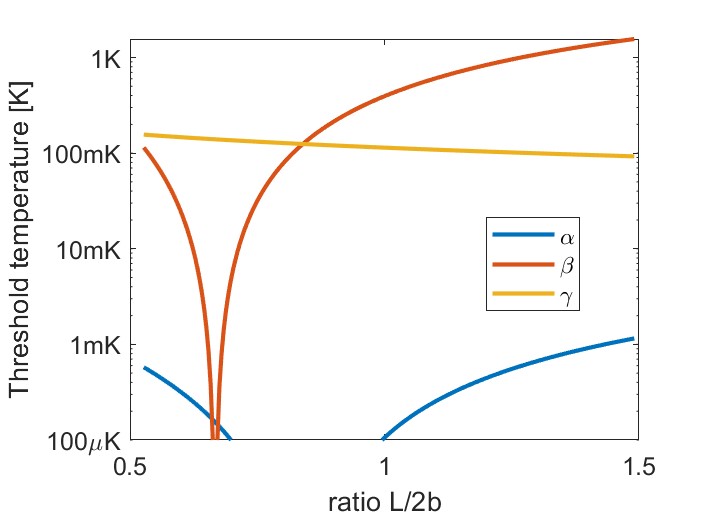}
\caption{Threshold temperatures [Eq.\,(\ref{eq:T_lib})] for libration dynamics for the three Euler angles of a ND in a Paul trap with parameters as in Fig.\,\ref{fig:frequencies_Lbratio}. At the transition point from a stable minimum to a metastable alignment, some values of the threshold temperature drop below the scale shown in the plot. In the case of an asymmetric shape (right sides of Fig.\,\ref{fig:Tth_vs_r} and this figure) cooling all the rotational modes to below 1\,mK ensures full angular trapping. In this work we consider rotation cooling only from the point where the angular DoF are already trapped in a librational mode.} 
\label{fig:Tth_vs_r_Lbratio}
\end{figure}

The above results for the secular frequencies provide an estimation of the temperatures below which the dynamics of a given angle in a thermal ensemble of NDs will be dominated by librational motion around the equilibrium alignment rather than rotational motion. These temperatures can be estimated as
\be T_i^{\rm lib}\sim \frac{1}{k_B}I_i(\omega^{(s)}_i)^2\phi_{\rm max}^2\,, \quad \phi_{\rm max}=\frac{\pi}{4} \,. 
\label{eq:T_lib} \ee 
As discussed after Eq.\,(\ref{eq:q}), if the charge to mass ratio, which determines the CoM trapping frequencies, is kept constant, then the libration frequencies depend only on the shape of the object and its orientation in the trap. It then follows that the threshold temperature for libration for a given shape scales linearly with the moment of inertia, namely, as $T^{lib}\propto M^{5/3}$ with the mass.

In Fig.\,\ref{fig:frequencies} we show the values of the libration frequencies of a cylindroidal ND (elliptical basis with semi-axes $a$ and $b$ and height $L$) in a Paul trap as a function of the ratio $b/a$ between the two diameters of the elliptic basis of the cylindroid. The quadrupole moments for a cylindroid object with uniform surface charge are given in Appendix\,\ref{app:Q}. The axes of rotation in the body frame are $\hat{\bf n}_1$ and $\hat{\bf n}_2$ along the elliptical basis axes and $\hat{\bf n}_3$ along the height. The Paul trap curvature parameters $\boldsymbol{\kappa}$ are $\kappa_z=2$ and $(\kappa_x,\kappa_y)=-(1-\epsilon,1+\epsilon)$ with $\epsilon=0.05$, giving rise to a slight asymmetry in the $x-y$ plane. This implies that if the ND is elongated along the $\hat{\bf n}_3$ axis (i.e., $L>2b\geq 2a$) then the equilibrium orientation of the ND is such that the $\hat{\bf n}_3$ axis is along the $x$ axis and the $\hat{\bf n}_2$ axis is along $y$ (if $b>a$). This corresponds to the Euler angles $\boldsymbol{\phi}=(0,\pi/2,0)^T$. In Fig.\,\ref{fig:frequencies} we show the numerical result for the libration frequencies around this equilibrium alignment when solving the full equations of motion in the Paul trap and compare these results to the approximate value of the libration frequencies obtained from the secular approximation [Eq.\,(\ref{eq:ws_approx})]. The agreement is excellent as long as the secular approximation is valid. In Fig.\,\ref{fig:frequencies_Lbratio} we show the secular libration frequencies for a ND with a fixed ratio $b/a=1.25$ as a function of the ratio $L/2b$ between the cylindroid height and the width along the wide axis of the elliptic basis. 

The threshold temperatures for transition from rotational dynamics to librational dynamics in the different angular modes are shown in Fig.\,\ref{fig:Tth_vs_r} for the same modes as in Fig.\,\ref{fig:frequencies} as a function of the ratio between the two axes of the elliptic cross-section of the cylindroid. In Fig.\,\ref{fig:Tth_vs_r_Lbratio} the threshold temperature is shown for the parameters appearing in Fig.\,\ref{fig:frequencies_Lbratio} as a function of the ratio between the cylindroid height and the largest length of the elliptic basis. These plots demonstrate the importance of the asymmetry of the ND shape and of the Paul trap potential in achieving an effective cooling of the angular DoF.  

\section{Parametric feedback cooling}
\label{sec:cooling}

\begin{figure}
\includegraphics[width=\columnwidth]{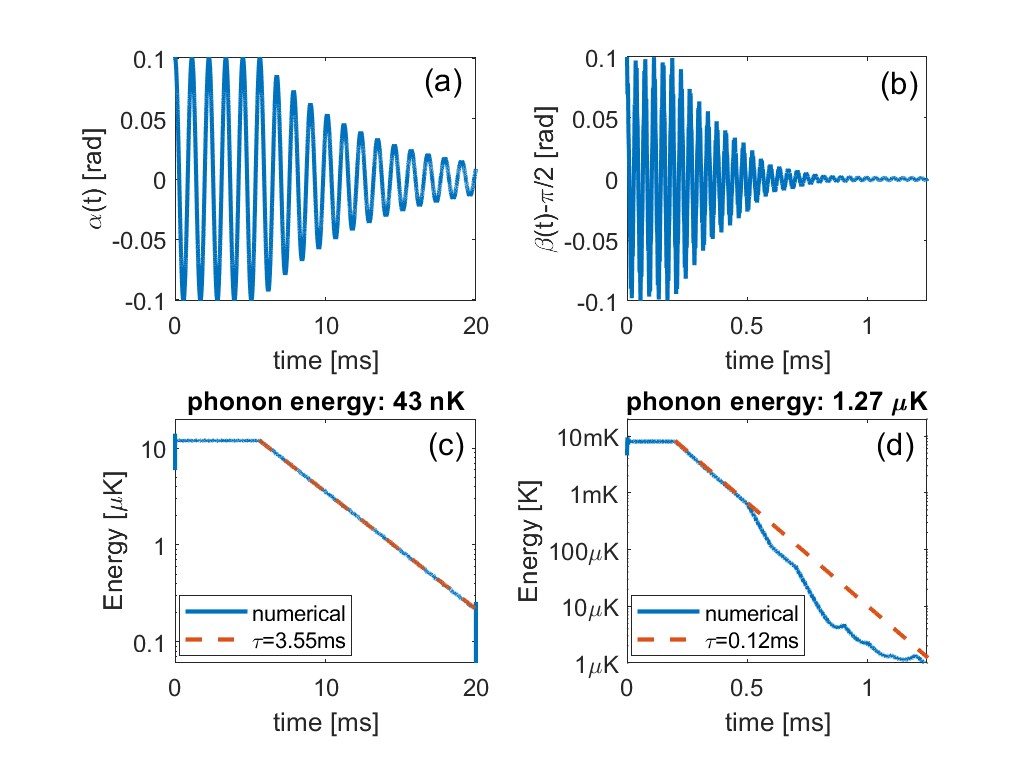}
\caption{Damping of librations by parametric feedback. A cylindrical ND ($a=b=30$\,nm and $L=100$\,nm) in a Paul trap with parameters as in Fig.\,\ref{fig:frequencies} is initially aligned 0.1\,rad from the equilibrium alignment ($\alpha=0$, $\beta=\pi/2$, long axis in the lowest gradient direction $x$). The frequency and phase of the oscillation is measured and obtained each 0.1\,ms, and the voltage on the electrodes is modulated with twice the frequency and relative amplitude of $\delta =0.05$. (a) and (b): damping of the oscillations in $\alpha$ and $\beta$, respectively. (c) and (d): damping of the energy $\frac12 I_i(\dot{\phi}_i^2+\omega_i^2\phi_i^2)$ (in units of Boltzmann's constant) for $\phi_i=\alpha$ and $\beta$, and $\omega_i$ being the respective secular frequencies. The secular frequency $\omega_{\beta}$ is larger than $\omega_{\alpha}$ by a factor of about 30, so that the corresponding initial energy is larger by three orders of magnitude for the mode $\beta$. The cooling starts after a minimal period that contains a few oscillations that can be detected.  The energy of the mode $\beta$ in (d) approaches the ground state energy after about 1\,ms of cooling, such that the classical calculation is not valid anymore. An exponential decay model with $\Gamma=\omega_i\delta$ (dashed curves) is in excellent agreement with the numerical results for $\alpha$ and a reasonable agreement for the $\beta$ mode. Here, the modes were assumed to be uncorrelated, i.e., when one mode is excited, the other modes are taken to be at zero temperature.}   
\label{fig:libration_damping}
\end{figure}

Feedback cooling of levitated objects is based on continuous weak measurement of the motion of the object and a feedback loop that induces a force that counteracts the instantaneous momentum but in the same time pushes the object to the potential minimum. If the motion is nearly oscillatory $x(t)\approx x_0\sin(\omega t+\phi_0)$, then this motion could be damped if the force pushes the object towards the center at each time when the object is moving away from the center. In a harmonic trap with frequency $\omega$, this damping may be achieved if the trap frequency is modulated with twice the frequency, $\omega^2\to \omega^2[1+\delta\sin(2\omega t+2\phi_0)]$, such that the feedback force is 
\begin{eqnarray} 
F^{\rm mod}(t) &=& -m\delta \omega^2x(t)\sin[2(\omega t+\phi_0)]\approx \nonumber \\
&& \approx -\delta \frac{2 m\omega}{x_0^2}x(t)^2\dot{x}(t)\,, 
\end{eqnarray}
where $\dot{x}(t)\approx \omega x_0\cos(\omega t+\phi_0)$ is the instantaneous velocity. 
For $\delta\ll 1$ the integrated decrease of the kinetic energy over one period gives
\begin{eqnarray} 
\Delta(\frac12mv^2)_{\rm period} &=& \int_0^T dt\,v(t)F^{\rm mod}(t)\approx \nonumber \\
&& \approx -m\omega^3x_0^2\delta \times \nonumber \\
&& \times \int_0^T dt\, \sin(2\omega t)\,\cos\omega t\sin\omega t= \nonumber \\
&& = -2\pi \delta\left(\frac12 mv_0^2\right)\,, 
\end{eqnarray}
where $v_0=\omega x_0$ is the initial velocity. It follows that the energy damping rate is 
\be \Gamma^{\rm damp}\approx \frac{2\pi \delta}{T}= \omega \delta\,. 
\label{eq:Gamma_damp} \ee
This is the principle of parametric feedback cooling in a harmonic oscillator potential. In order to implement this procedure, the phase $\phi(t)=\omega t+\phi_0$ needs to be continuously measured. 
The efficiency of the cooling depends on the ability to accurately measure the phase of the motion during the evolution. 

If we also consider heating mechanisms, such as collisions with thermal background gas molecules and interaction with the light beam used for detection then the effective differential equation for the energy in a given mode of oscillation may be written as
\be \frac{d\bar{E}}{dt}=-\Gamma^{\rm damp}\bar{E}+\left(\frac{d\bar{E}}{dt}\right)_{\rm heating}\,. \ee
The cooling + heating process would reach a steady state at a final temperature
\be T_{\rm s-s}=\frac{1}{k_B\Gamma^{\rm damp}}\left(\frac{d\bar{E}}{dt}\right)_{\rm heating}\,.
\label{eq:Tss}  \ee

In what follows, we demonstrate the process of parametric feedback cooling for the rotational DoF in a Paul trap and estimate the minimal temperature based on the expected damping rate and heating rates.
Unlike the translational motion, which may be harmonic under the secular approximation, the dynamics of the rotational modes is inherently nonlinear and becomes harmonic only for small angular deviations from the equilibrium alignment. It follows that the cooling of this motion might be more challenging than the cooling of the translational motion. 

Fig.\,\ref{fig:libration_damping} demonstrates the damping of single libration modes due to parametric feedback cooling, where the oscillation signal is measured and analyzed each 0.1\,ms. The frequency and phase of the oscillation is retrieved by spectral analysis of a signal accumulated over a suitable time duration and then applying a band-pass filter to isolate the specific oscillation, followed by a fit of the filtered signal to a cosine function. The procedure starts after a time duration that allows taking the signal over a suitable time duration, which must contain at least a few oscillation periods. As shown in the figure, damping of a fast oscillation (in this specific case, libration of the angle $\beta$ around the $\hat{\bf n}_2$ axis) is much faster than damping of a slow libration (in this case, oscillation of the $\alpha$ angle along the $\hat{\bf n}_1$ axis). In this example the shape of the object was assumed to be cylindrical so that damping of rotations along the $\hat{\bf n}_3$ axis (angle $\gamma$) is not possible by using the electric potential. Note that in both simulations, we have ignored the dynamics around other axes when damping the librations around a given axis. For both cases, we obtain a reasonable (for mode $\beta$) or even excellent (for mode $\alpha$) agreement with the exponential damping law with the damping rate given by Eq.\,(\ref{eq:Gamma_damp}). 

\begin{figure}
\includegraphics[width=\columnwidth]{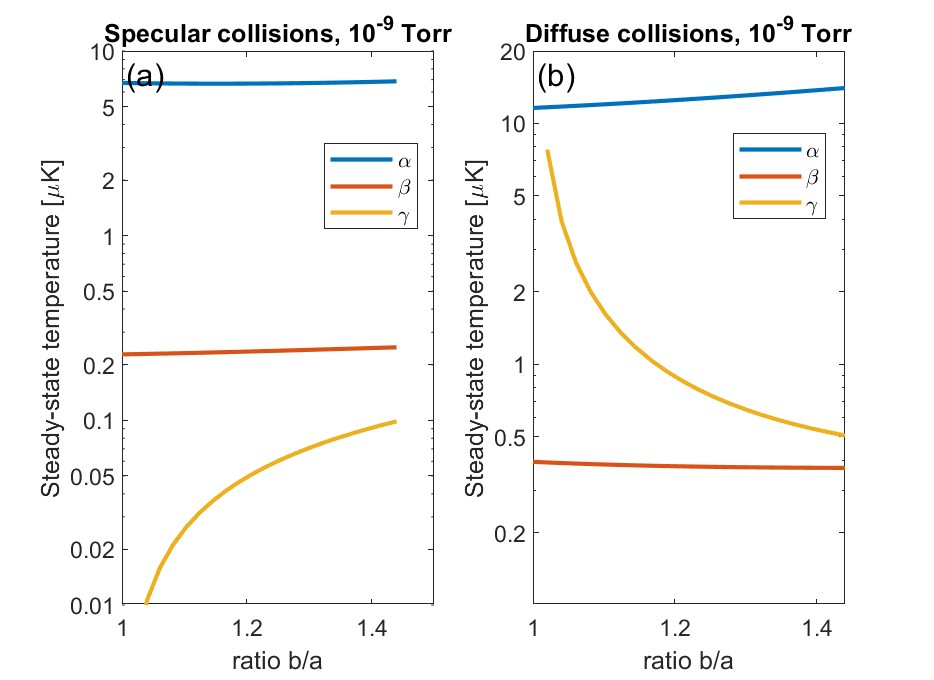}
\caption{Steady-state temperatures (minimal temperatures) for the librational modes of a ND after ideal parametric feedback cooling in a $10^{-9}$\,Torr vacuum level at room temperature ($T=300$K). The ND parameters are those of Fig.\,\ref{fig:frequencies} and the cooling rates were taken to be proportional to the libration frequency [Eq.\,(\ref{eq:Gamma_damp})] with modulation depth $\delta=0.05$. 
Plots (a) and (b) show two models for the gas collisions: specular collisions at the surface of the ND or diffuse collisions (see Appendix\,\ref{app:Gamma_gas}). 
The minimal temperature for this heating mechanism is linearly proportional to the background gas pressure, so that the results should be multiplied by the pressure factor to obtain minimal temperatures for different pressures.} 
\label{fig:Tss_vs_r}
\end{figure}

The heating mechanism that we consider here is collisions with background gas molecules (specifically air molecules). The heating rate of a mode $j$ is given by
\be \left(\frac{d\bar{E}_j}{dt}\right)_{\rm gas}=\frac12k_BT_{\rm gas}\Gamma_j^{\rm coll}\,, 
\label{eq:heating_gas}\ee \
where $\Gamma_i^{\rm coll}$ depends on the pressure and temperature of the gas molecules, as well as on the shape of the object, the mass of the molecules and the nature of the collisions. Here we have taken the expression for $\Gamma_j^{\rm coll}$ from Eq.\,(40) in Ref.\,\cite{Martinetz2018} and calculated them for the specific ND shape, which we have chosen to be cylindroidal (see calculation in Appendix\,\ref{app:Gamma_gas}). 
It follows from Eqs.\,(\ref{eq:Tss}) and\,(\ref{eq:heating_gas}) that the steady-state temperature of a given mode is 
\be T_j^{\rm s-s}=\frac{\Gamma_j^{\rm gas}}{\Gamma_j^{\rm gas}+\Gamma_j^{\rm damp}}T_{\rm gas}\,. \ee

Fig.\,\ref{fig:Tss_vs_r} shows the steady-state temperature (minimal temperature) of the librational modes of a ND for a process that consists of parametric feedback cooling [with cooling rates given by Eq.\,(\ref{eq:Gamma_damp})] and collisions with gas molecules. While the rotation/libration mode related to the Euler angle $\gamma$ is more difficult to cool when the object is close to symmetric around the $\hat{\bf n}_3$ axis, the minimal temperature that this mode can reach is quite low because the collisions with the gas molecules are also not effective in heating. On the other hand, it is much more difficult to reach a low temperature for the mode $\alpha$ (rotation in the $x-y$ plane) because the collisions with gas molecules are efficient in heating, while the weak trapping in this direction does not allow a fast cooling rate by the electric forces. 

The nanoparticle orientational motion can be monitored in real-time
by measuring the amplitude of the scattered optical field. One possibility
is to use an optical field propagating perpendicular to the long axis of the particle (the positive y-axis in Fig.\,\ref{fig:setup}) with the polarization along the long axis of the particle (vertical x-axis in Fig.~\ref{fig:setup}).
Here we consider the regime of low-intensity optical fields such that the radiation pressure force on the trapped nanoparticle remains negligible (and similarly, we suppose that the optical gradient trap is negligible
compared to the Paul trap, which is used to trap the orientational DoFs). Alternatively, one can also consider configurations with two, equal-intensity and counter-propagating beams, such that
the radiation pressure force cancels out exactly. 

The time-trace signal, extracted from the measured optical field, will contain the combined information about all three orientational degrees of freedom (as well as in general of the translational DoFs). For simplicity of illustrating the order of magnitude of the noises, let us here suppose that due to the separation in frequency
space we can isolate a single orientational degree of freedom, e.g., the $\alpha$ motion. We suppose that the orientational degree of freedom is librating such that we can adopt the familiar modelling of harmonic oscillators: the on resonance power spectral density is proportional to $S_{\text{lib}}\propto2k_{B} \Gamma_\text{gas} T_{\text{gas}}/((\Gamma_\text{damp})^2 I\omega{}^{2})$, where $T_{\text{gas}}$ is the gas temperature, $I\equiv I_{1}$ is the appropriate moment of inertia for $\alpha$ librations, $\omega\equiv \omega_1$ is the frequency of the corresponding librational mode, $\Gamma_\text{gas}=\Gamma^\text{gas}_1$ is the damping rate related to gas collisions, and $\Gamma_\text{damp}\equiv\Gamma^\text{damp}_{1}$ is the effective cooling rate introduced earlier (we assume $ \Gamma^\text{damp}_{1}\gg \Gamma^\text{gas}_{1}$).
Here we have neglected the back-action noise from the detection as well as the recoil heating from photon scattering as it should remain negligible at the considered low optical intensities (in comparison to the noise related to gas collisions). The imprecision noise floor can be however estimated from the orientational photon recoil heating terms. We first make the approximation $\alpha\approx0$ and $\beta\approx\pi/2$, which is suitable for small librations (i.e, we assume the system is cooled such that the angular oscillations are small). For simplicity and concreteness, we also assume that the electric susceptibility along the long axis, $\chi\equiv\chi_3$, is much larger than the  electric susceptibilities along the other two orthogonal axis, $\chi_1$ and $\chi_2$, such that we can make the approximation $\vert \chi_3-\chi_2\vert \approx \chi_3$ (and similarly $\vert\chi_3-\chi_1\vert\approx \chi_3$).
We then find a simple expression for the imprecision noise floor, $S_{\text{imp}}\propto(\eta(\sigma_{R}/\sigma_{L})(P/(\hbar\omega_{L}))^{-1}$,
where $\eta$ is the detection efficiency, $\sigma_{R}=\sigma_{R}(\chi)$ is the Rayleigh
cross-section area, $\sigma_{L}$ is the effective beam cross-section area, $P$ is the laser power, $\omega_{L}=2\pi c/\lambda$, and $\lambda$ is the optical wavelength. From the experimental numbers considered in this work we find that the signal to noise ratio, estimated as $\sim S_{\text{lib}}/S_{\text{imp}}$, does not pose a fundamental
limitation for reasonable values of the laser power and of the detection efficiency~\citep{vinante2019testing}. However, other technical sources of noise, in particular, the source of dark noise present in the detector, will present a limiting factor in the detection of the orientational motion.

\section{Discussion and outlook}
\label{sec:discussion}

In this work, we have studied trapping and cooling of the rotational/librational DoF of a charged ND in a Paul trap. We have shown in Sec.\,\ref{sec:motivation} that for performing SG interferometry in the $T^3$ phase scheme, it is essential to cool the rotational DoF to a few hundred rotational phonons. We have proposed a parametric feedback cooling procedure and showed that in a vacuum of about $10^{-9}$\,Torr, the required temperatures should be within reach in the near future. We clearly demonstrate that libration modes that are more tightly trapped (i.e., have higher libration frequencies) are cooled faster and can reach a lower temperature (in absolute terms and even more in terms of number of phonons). For efficient cooling it is therefore advantageous to reach higher trapping frequencies of the Paul trap (see our work on a high-frequency Paul trap\,\cite{Skakunenko_Needle_trap}) and larger asymmetry of the trapping field. 

In order to estimate the practical limits on cooling, it is necessary to take into account additional factors that determine the efficiency and accuracy of the cooling process. These factors include the signal-to-noise ratio of the detection, which allows the feedback, the accuracy of the phase of the modulations of the electric fields, and other possible sources of noise, such as optical noise from the detection light or black-body radiation\,\cite{Martinetz2022}. 

In addition, we need to investigate more thoroughly the effects of cross-talk between the different translational and rotational/librational modes during the cooling process. It is essential to examine the way to cool the rotational DoF from the point where they are in a free rotational state. This cooling may involve sympathetic cooling of a few modes during the cooling of other modes that are easier to cool. 

Finally, it may very well be that the requirement for rotation cooling will be considerably relaxed if one uses gyroscopic stabilization. However, we believe that it will be necessary to study the proposed procedure of gyroscopic stabilization beyond the existing studies that were published recently\,\cite{ZhouT2024,Wachter2025,Rizaldy2025}. This study should take into account the cooling mechanisms and the specific interferometric scenario that will enable using this method for overcoming the limitations of the uncertainty principle for the rotational modes. 

\begin{acknowledgments}
This work was funded by the Gordon and Betty
Moore Foundation (doi.org/10.37807/GBMF11936), and Simons Foundation
(MP-TMPS-00005477). MT acknowledges funding from the Slovenian Research and Innovation Agency (ARIS) under contracts N1-0392, P1-0416, SN-ZRD/22-27/0510.
MM thanks the Ariane de Rothschild Women Doctoral Program for outstanding female PhD students for its support.
\end{acknowledgments}


\appendix

\section{Calculation of quadrupole charge moments for a few shapes}
\label{app:Q}

\subsection{Box}
The quadrupole moments of a rectangular box with half side-lengths $(a,b,c)$ with uniform surface charge density is
\begin{eqnarray}
Q_{11} &=& q\frac{ab+ac+3bc}{ab+ac+bc}\frac{a^2}{3}\,, \\
Q_{22} &=& q\frac{ab+3ac+bc}{ab+ac+bc}\frac{b^2}{3}\,, \\
Q_{33} &=& q\frac{3ab+ac+bc}{ab+ac+bc}\frac{c^2}{3}\,. 
\end{eqnarray}

The moments of inertia are
\be I_1=\frac{M}{3}(b^2+c^2)\,, \quad I_2=\frac{M}{3}(a^2+c^2)\,, \quad I_3=\frac{M}{3}(a^2+b^2)\,. \ee

\subsection{Cylindroid}

Consider an object having a shape of a cylinder with ellipsoidal base having semi-axes $a$ and $b$ and height $L$. The charge is assumed to be uniformly distributed on the surface, such that $\rho_S=q/[2\pi ab+\pi(a+b)A_0(a/b) L]$, where $A_0(r)=[4/\pi(1+r)]\mathrm{CE}_i(r^2)$ with $\mathrm{CE}_i$ being the complementary complete elliptic integral of the second kind. For $0<r<\infty$ the function $A_0(r)$ ranges between 1 (at $r=1$) and 1.2732 (at $r\to 0$ or $r\to \infty$). 
The quadrupole moments of the charge are
\begin{eqnarray}
Q_{11} &=& \rho_S\left[La^2\int_0^{2\pi}d\phi\,\sqrt{a^2\sin^2\phi+b^2\cos^2\phi}\cos^2\phi \right. +\nonumber \\
&& \left.+2\int_{-b}^b dy\int_{a\sqrt{1-y^2/b^2}}^{a\sqrt{1-y^2/b^2}}dx\,x^2\right] = \nonumber \\
&=& \frac{\pi}{2}a^2\rho_S\left[L(a+b)A_1(b/a) + ab\right]\,, 
\label{eq:Q11}  \\ 
Q_{22} &=& \frac{\pi}{2}b^2\rho_S\left[L (a+b)A_1(a/b) + ab\right]\,, 
\label{eq:Q22} \\ 
Q_{33} &=& \frac{\pi}{2}L^2\rho_S\left(\frac{1}{6} (a+b)A_0(a/b) L+ab\right)\,. 
\label{eq:Q33}  
\end{eqnarray}
Here 
\be A_1(r) = \frac{2}{\pi(1+r)}\int_0^{2\pi}d\phi\,\sqrt{\sin^2\phi+r^2\cos^2\phi}\cos^2\phi\,d\phi\,, \ee
such that $0.79<A_1(r)<1.7$.
For a cylinder ($a=b$) $A_0=A_1=1$ and $\rho_S=q/[2\pi a(a+L)]$. We obtain the quadrupole moments
\begin{eqnarray}
Q_{11} &=& Q_{22}=\frac{qa^2}{2}\frac{L+\frac12 a}{L+a}\,, \\ 
Q_{33} &=& \frac{qL^2}{4}\frac{a+\frac13 L}{a+L}\,. 
\end{eqnarray}

For calculating the moments of inertia of this shape, we need to perform the volume integrals $\langle x^2\rangle$, $\langle y^2\rangle$ and $\langle z^2\rangle$ over the shape. We then obtain
\be \frac{\rho_M}{M} \langle x^2\rangle=\frac14a^2\,, \quad \frac{\rho_M}{M}\langle y^2\rangle=\frac14b^2\,, \quad \frac{\rho_M}{M}\langle z^2\rangle=\frac{1}{12}L^2\,. \ee
The moments of inertia are then
\begin{eqnarray}
I_1 &=& \rho_M(\langle y^2\rangle+\langle z^2\rangle)=\frac14 M(b^2+\frac13 L^2)\,, \\
I_2 &=& \frac14 M(a^2+\frac13L^2)\,, \\
I_3 &=& \frac14M(a^2+b^2)\,. 
\end{eqnarray}

\section{Damping rates due to background gas collisions}
\label{app:Gamma_gas}

Here we calculate the damping rates due to background gas collisions for the rotational modes of a ND having a cylindroid shape. The equation for the damping rates is taken from Ref.\,\cite{Martinetz2018}. The damping tensor is given by

\begin{multline}
    \boldsymbol{\Gamma}_{\text{rot}} =
    n_g \sqrt{\frac{m_g k_B T}{2\pi}} \cdot \\
    \cdot \int_{\partial V} dA \bigg[
    \left(4 - 3\alpha_c + \frac{\pi}{2} \alpha_c \gamma_s\right)
    (\mathbf{r} \times \hat{\mathbf{n}}) \otimes (\mathbf{r} \times \hat{\mathbf{n}})
    +\\
    +\alpha_c(r^2\mathbb{1}-\mathbf{r}\otimes \mathbf{r})
    \bigg] \cdot A_2
\mathbf{I}^{-1}
\end{multline}

where
$n_g$ is the gas number density: $n_g = \frac{P}{k_B T}$,
$m_g$ is the gas molecule mass,
$k_B$ is the Boltzmann constant,
$T$ is the background gas temperature (K),
$\alpha_c$ is the accommodation coefficient (1 - fully diffuse, 0 - fully specular),
$\gamma_s = \sqrt{T_s / T}$  is the ratio of surface to gas temperature,
The vector $\mathbf{r}$ denotes the position of a surface element relative to the CoM,
$\hat{\mathbf{n}}$ is the normal vector at $\mathbf{r}$,
$dA$ is the local surface area element,
$\partial V$ is the surface of the body,
$\mathbf{I}^{-1}$ is the inverse of the moment of inertia tensor (in the body frame).

Here we assume a symmetric body, so that the $\boldsymbol{\Gamma}$ tensor is diagonal. 
By parameterizing the vectors along the cylindroid height in terms of the phase $\phi$, we have
\be \mathbf{r}=\left(\begin{array}{c} a\cos\phi \\ b\sin\phi \\ z\end{array}\right)\,, \ee 
\be \hat{\mathbf{n}}=\frac{1}{\sqrt{a^2\sin^2\phi+b^2\cos^2\phi}}\left(\begin{array}{c} b\cos\phi \\ a\sin\phi \\ 0\end{array}\right)\,. \ee   
The surface element is $dA=\sqrt{a^2\sin^2\phi+b^2\cos^2\phi}d\phi\,dz$. We therefore obtain
\begin{align}
\Gamma_{11} &= \frac{n_gp_g(T)}{I_1}\left[\eta \left(\frac{L^3}{6}aA_2\left(\frac{b}{a}\right)+ ab^3\right)+\alpha_c(\bar{Q}_{22}+\bar{Q}_{33})\right]\,, \\ 
\Gamma_{22} &= \frac{\pi}{2}\frac{n_gp_g(T)}{I_2}\left[\eta\left(
\frac{L^3}{6}bA_2\left(\frac{1}{b}\right)+a^3b\right)+\alpha_c(\bar{Q}_{11}+\bar{Q}_{33})\right]\,, \\ 
\Gamma_{33} &= \frac{\pi}{2}\frac{n_gp_g(T)}{I_3}\left[\eta \frac{L}{2a}(a^2-b^2)^2 A_3\left(\frac{b}{a}\right)
+\alpha_c(\bar{Q}_{11}+\bar{Q}_{22})\right]\,, 
\end{align}

where $ p_g=\sqrt{m_gk_BT/2\pi}$, $\eta=\frac{\pi}{2}(4 - 3\alpha_c + \frac{\pi}{2} \alpha_c \gamma_s)$, $\bar{Q}_{jj}=Q_{jj}/\rho_S$ [Eqs.\,(\ref{eq:Q11})-(\ref{eq:Q33})]and
\be A_2(r)=\frac{1}{\pi}\int_0^{2\pi}\frac{\sin^2\phi\,d\phi}{\sqrt{\sin^2\phi+r^2\cos^2\phi}}\,, \ee
\be A_3(r)=\frac{1}{\pi}\int_0^{2\pi}\frac{\sin^2(2\phi)\,d\phi}{\sqrt{\sin^2\phi+r^2\cos^2\phi}}\,, \ee

%

\end{document}